\documentclass[a4paper, twocolumn, twoside,
superscriptaddress,
 amsmath,amssymb,
pra,
]{revtex4-2}

\usepackage{mathtools}
\usepackage{amsmath,amssymb, amsthm, amsfonts}
\usepackage{dsfont} 
\usepackage{xspace}
\usepackage{xcolor}
\usepackage{changes}
\usepackage{verbatim}
\usepackage{float}
\usepackage{comment}
\usepackage{soul}

\usepackage{multirow}
\usepackage{color}
\usepackage{graphicx}

 \usepackage[unicode=true, breaklinks=false, pdfborder={0 0 1}, backref=false, colorlinks=true, linkcolor=blue, citecolor=blue, urlcolor=blue]{hyperref}
\newcommand{\expval}[1]{\left\langle #1 \right\rangle}

\DeclareMathAlphabet{\mathbbold}{U}{bbold}{m}{n}

\newcommand{\bb}[0]{\begin{eqnarray}}
\newcommand{\ee}[0]{\end{eqnarray}}

\begin{document}

\title{
Current-based metrology with two-terminal mesoscopic conductors}

\author{Shishir Khandelwal}
\email{shishir.khandelwal@fysik.lu.se}
\affiliation{%
Department of Applied Physics, University of Geneva, 
CH-1211 Geneva, Switzerland
}%
\affiliation{Physics Department and NanoLund, Lund University, Box 118, 22100 Lund, Sweden}

\author{Gabriel T. Landi}
\email{gabriel.landi@rochester.edu}
\affiliation{Department of Physics and Astronomy, University of Rochester, Rochester, New York 14627, USA}
\affiliation{University of Rochester Center for Coherence and Quantum Science, Rochester, New York 14627, USA}

\author{Géraldine Haack}
\email{geraldine.haack@unige.ch}
\affiliation{%
Department of Applied Physics, University of Geneva, 
CH-1211 Geneva, Switzerland
}

\author{Mark T. Mitchison}
\email{mark.mitchison@kcl.ac.uk}
\affiliation{School of Physics, Trinity College Dublin, College Green, Dublin 2, D02 K8N4, Ireland}
\affiliation{Department of Physics, King’s College London, Strand, London, WC2R 2LS, United Kingdom}

\date{\today}

\begin{abstract}
The traditional approach to quantum parameter estimation focuses on the quantum state, deriving fundamental bounds on precision through the quantum Fisher information. In most experimental settings, however, performing arbitrary quantum measurements is highly unfeasible. In open quantum systems, an alternative approach to metrology involves the measurement of stochastic currents flowing from the system to its environment. However, the present understanding of current-based metrology is mostly limited to Markovian master equations. Considering a parameter estimation problem in a two-terminal mesoscopic conductor, we identify the key elements that determine estimation precision within the Landauer-Büttiker formalism. Crucially, this approach allows us to address arbitrary coupling and temperature regimes. Furthermore, we obtain analytical results for the precision in linear-response and zero-temperature regimes.  For the specific parameter estimation task that we consider, we demonstrate that the boxcar transmission function is optimal for current-based metrology in all parameter regimes.   

\end{abstract}

\maketitle

\textit{Introduction.---} Quantum parameter estimation is one of the cornerstones of quantum technologies \cite{Giovannetti2011} and has been the subject of a wide range of theoretical \cite{Alipour2014,Mehboudi2019,Mitchison2020,PerarnauLlobet2020,PerarnauLlobet2021,Salvia2023,Abiuso2024} and experimental \cite{Wang2019,Liu2021,Marciniak2022,Long2022,Yin2023,Valeri2023} works in recent years. A considerable portion of theoretical research in this area focuses on the quantum state \cite{Paris2009,Toth2014}, through which key statistical quantities such as the quantum Fisher information can be determined. Importantly, this approach sheds light on the fundamental bounds on precision and optimal measurements for quantum metrology. However, achieving optimal precision can require heavily engineered and experimentally unfeasible quantum measurements. This becomes even more challenging in open quantum systems, where environmental noise can reduce the achievable precision and further complicate the identification and implementation of the optimal measurement~\cite{Haase2016}.

In open systems, the measurement of quantum transport observables such as currents is often more feasible. This has motivated systematic statistical analyses of metrology using the output currents of open quantum systems~\cite{Landi2023}. There are several crucial questions: for example, to determine the key quantities of interest in current-based metrology and to identify the systems and transport conditions that allow for achieving better precision. A strong focus has been on parameter estimation using continuous monitoring of open quantum systems \cite{Mabuchi1996}, which can often be interpreted as photocurrent-based metrology \cite{Gambetta2001,Stockton2004,Tsang2011, Ralph2011, Tsang2013, Gammelmark2013,Kiilerich2014,Catana2015,Macieszczak2016, Ng2016, Albarelli2017,Albarelli2018,Clark2019,Rossi2020,Ilias2022, Tsang2023, Yang2023,Godley2023,Cabot2024}.  However, most previous works have focussed primarily on optical systems, where the environment can usually be treated within a weak-coupling, Markovian approximation: an assumption that is generally not satisfied in low-temperature transport experiments involving mesoscopic electronic devices. 

\begin{figure}
    \centering
    \includegraphics[width=1\columnwidth]{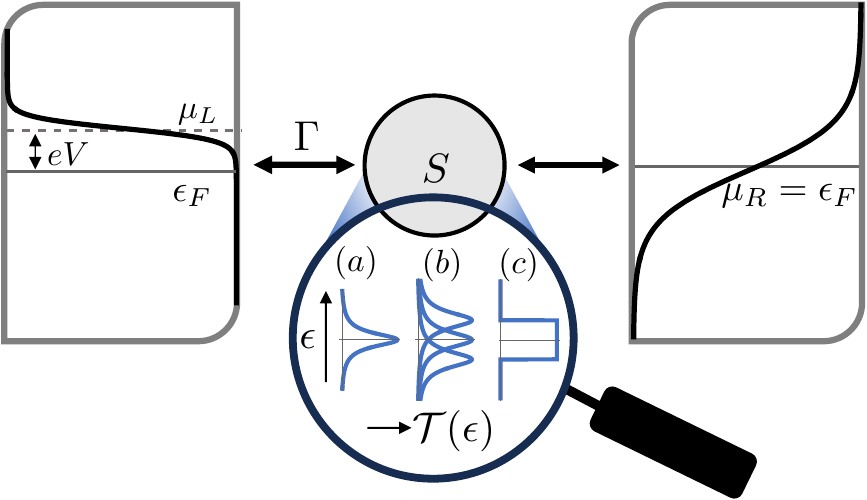}
    \caption{A two-terminal mesoscopic conductor embedded between two fermionic leads. The energy bias $eV$ provided by the bias voltage pushes the system out of equilibrium. The transport properties depend on the transmission function $\mathcal T(\epsilon)$; we consider (a) Lorentzian, (b) sums of Lorentzians and (c) boxcar shapes. }
    \label{fig:setup}
\end{figure}

In this work, we address the problem of parameter estimation in two-terminal quantum conductors. In these systems, the electrical current is determined by both the properties of the system and the transport conditions. Therefore, current measurements represent a natural path for parameter estimation. We note that Mihailescu et al.~\cite{Mihailescu2024} recently addressed quantum parameter estimation in mesoscopic electronic setups within an adiabatic linear-response formalism. Here, we focus on precision optimization within the Landauer-Büttiker framework, which holds arbitrarily far from equilibrium so long as interactions between electrons can be neglected (or treated within a mean-field approximation). This approach allows us to determine the optimal conditions for steady-state metrology under arbitrary coupling, voltage bias, and temperature regimes. In this setting, within the linear-response and zero-temperature regimes, we obtain analytical bounds for the precision, identifying and physically interpreting the fundamental elements that determine precision in current-based metrology. We also demonstrate numerically that the optimal transmission function for current-based metrology is the boxcar function (i.e., energy-independent transmission within a given window) (see Fig. \ref{fig:setup}), both within and far beyond the linear-response regime.

\textit{Current-based metrology.---} 
\begin{figure*}
    \centering
    \includegraphics[width=\textwidth]{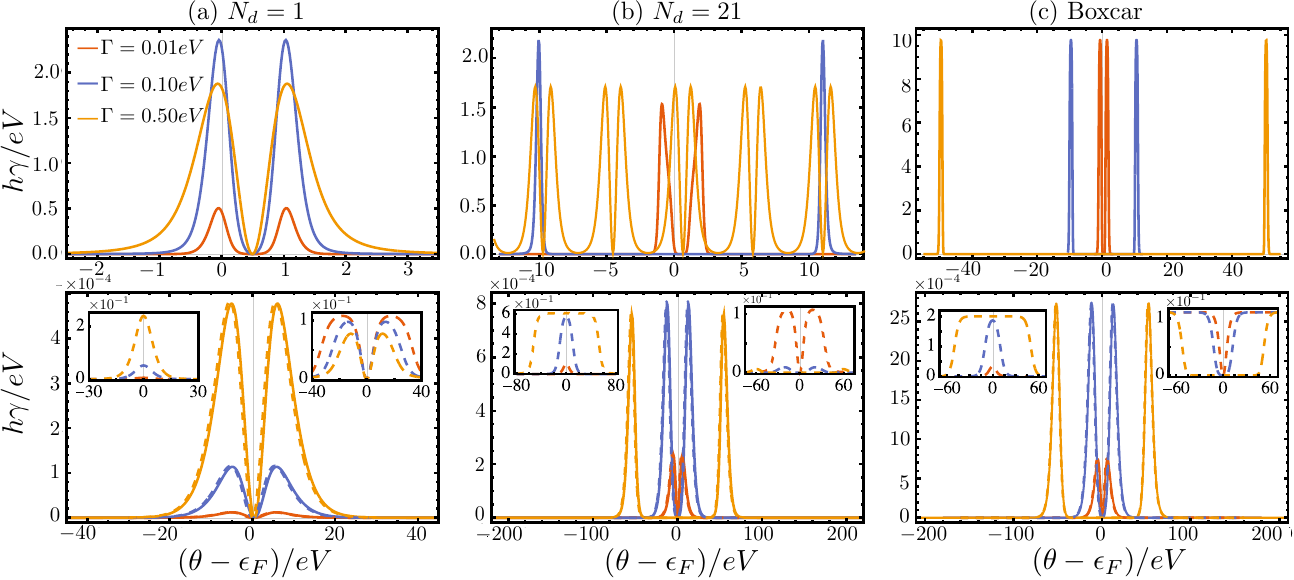}
    \caption{Precision rate $\gamma$ as a function of $\theta$, (a) single Lorentzian, $N=1$, (b) $N=21$ and (c) boxcar transmission function for $k_BT=0.1\,eV$ (top) and $k_BT=3\,eV$ (bottom), and three values of $\Gamma=0.01,0.1,0.5$ $eV$. Solid curves have been drawn with full Landauer-Büttiker expressions and dashed with the linear-response expression Eq. \eqref{eq:preclin}. In all bottom panels, the left inset shows the conductance as a function of $\theta$ and the right, relative sensitivity. $\delta=100\Gamma$ is set throughout the figure. }
    \label{fig:linrep}
\end{figure*}
We consider the two-terminal setup shown in Fig. \ref{fig:setup}, with a device embedded between two fermionic reservoirs indexed by $\alpha=L,R$.  As the mesoscopic device, we consider a two-terminal conductor which, within a scattering approach to transport, is fully characterized by its energy-dependent transmission function $\mathcal T(\epsilon)$. 
The reservoir modes are characterized by their  
temperatures $T_\alpha$ and chemical potentials $\mu_\alpha$. We consider thermal equilibrium corresponding to $T_L = T_R \equiv T$ and a finite bias voltage $V$. The energy bias between the two contacts is given by $eV$ with respect to the Fermi energy of the two contacts, $\epsilon_F$, i.e.~$\mu_L = \epsilon_F + e V$ and $\mu_R = \epsilon_F\,$. This bias sets up a steady-state current, $I$, flowing through the system. 

Our goal is to establish how well a current measurement can be used to infer some parameter, $\theta$, of the underlying system. In principle, $\theta$ may be a property of the reservoirs (for example, the temperature) or the system (for example, the energy of a quantum dot). We assume the current is measured by integrating the signal from an ammeter over a time window $\tau$. This is equivalent to measuring the charge, $Q$, transferred into one of the reservoirs (say, $\alpha=R$), which can be modelled by a two-point measurement of the electron number in the same reservoir~\cite{esposito_nonequilibrium_2009}. At long times, the mean $\langle Q\rangle$ and variance ${\rm Var}[Q]$ are related to the statistics of the current as $\langle Q\rangle = \tau \langle I \rangle$ and ${\rm Var}[Q] = \tau \langle \langle I^2 \rangle \rangle$, where $\langle I\rangle$ is the mean steady-state current and $\langle \langle I^2 \rangle \rangle$ is the DC current noise or diffusion coefficient. As shown in Appendix~\ref{app:charge_statistics}, these relations hold for $\tau \gg t_{\rm rel}, t_{\rm cor}$, where $t_{\rm rel}$ is the relaxation time to the steady state and $t_{\rm cor}$ is the decay time of the current-current autocorrelation function. 

We focus on local parameter estimation, where the unknown parameter $\theta$ lies close to some known value $\theta_0$, and assume that the mean charge $\langle Q\rangle = \Phi(\theta)$ is a known, differentiable function of $\theta$ in the vicinity of $\theta_0$. Then, it is easy to check that the estimator $\check{\theta} = \theta_0 + [Q - \Phi(\theta_0)]/\Phi'(\theta_0)$ is locally unbiased, i.e.~$\langle \check{\theta}\rangle = \theta + \mathcal{O}((\theta - \theta_0)^2)$, and has a variance given by ${\rm Var}[\check{\theta}] = {\rm Var}[Q]/(\partial_\theta \langle Q\rangle)^2 = (\gamma \tau)^{-1}$, where the \textit{precision rate} is defined by
\begin{equation}
    \label{eq:prec}
    \gamma = \frac{(\partial_\theta\langle I\rangle)^2}{\langle \langle I^2\rangle \rangle}.
\end{equation}
That is, the estimation error falls linearly in the time of charge  measurement, with a rate given by Eq.~\eqref{eq:prec}. As discussed in Appendix~\ref{app:Fisher}, $\gamma$ is upper bounded by the Fisher-information rate and this bound is saturated whenever the distribution of $Q$ is Gaussian, as expected at long times by the central-limit theorem~\footnote{In fact, Eq.~\eqref{eq:prec} can also be derived from a saddle-point approximation of the charge characteristic function as $\tau\to\infty$ (e.g.~see Sec.~V~E of Ref.~\cite{Landi2023}), without any explicit Gaussian assumption. We thank Patrick P. Potts for pointing this out to us.}. Both $\langle I\rangle$ and $\langle \langle I^2\rangle \rangle$ have exact expressions within Landauer-B\"uttiker theory, which are quoted explicitly in Appendix~\ref{app:B}. In the following, we use these expressions to analyse the estimation precision of current-based metrology in two-terminal devices.

\textit{Linear response at thermal equilibrium.---} At thermal equilibrium, in the linear-response regime for the charge current, $eV \ll  k_B T, \epsilon_F$, the current and noise are set by the conductance $G=-2\frac{e^2}{h} \int d\epsilon  \mathcal T(\epsilon) \partial_\epsilon f(\epsilon,T)$, $\langle I \rangle_{\text{LR}} = G \, V$
and $\langle \langle I^2 \rangle \rangle_{\text{LR}} = 4 k_BT \,G \,$ \cite{Blanter2000}. 
The precision rate $\gamma$ from Eq. \eqref{eq:prec} can therefore be expressed as
\begin{eqnarray}\label{eq:preclin}
    \gamma_{\text{LR}} = \frac{V^2 \, (\partial_\theta G)^2}{4 k_BT G} = \frac{1}{4}\frac{e V}{k_BT}\left( \partial_\theta \ln G \right)^2\frac{G \, V}{e}  \,.
\end{eqnarray}
This result enables a clear interpretation of the sensitivity of a two-terminal mesoscopic conductor. The first factor on the right-hand side is a dimensionless ratio that characterises the (inverse) scale of thermal fluctuations set by the energy transport window $eV$ and the thermal energy $k_B T$. The second factor captures the relative sensitivity to the parameter $\theta$, represented by the logarithmic derivative of the conductance, which is a property of the conductor. Finally, the last factor $GV/e$ corresponds to the particle current, i.e., the rate at which the charge is measured. 
 Equation \eqref{eq:preclin} is valid both when the parameter to be estimated is a property of the conductor (for example, the bare energy of a quantum dot) or a property of the reservoirs (for example, the temperature).

\textit{Zero-temperature limit.--} At zero temperature, the Fermi distributions behave as Heaviside functions, $f_\alpha(\epsilon) \equiv \Theta(\mu_\alpha - \epsilon)$. This sets a finite energy window of width $eV$ through which charge transport takes place. In general, this limit entails full (non-linear) response in the bias voltage. Therefore, full Landauer-Büttiker expressions for current and noise must be considered to calculate the precision. As shown in Appendix~\ref{app:B}, the zero-temperature precision then takes the form,
\begin{eqnarray}\label{eq:zerT}
    \gamma_{0} = \frac{1}{h} \frac{\left(\int_{\epsilon_F}^{\epsilon_F + eV}  d\epsilon \partial_\theta \mathcal T(\epsilon)\right)^2}{\int_{\epsilon_F}^{\epsilon_F + eV}  d\epsilon \mathcal T(\epsilon)(1-\mathcal T(\epsilon))}\,.
\end{eqnarray}Naturally, to have a non-trivial parameter estimation, $\theta$ must now be a property of the conductor, entering its transmission function $\mathcal{T}(\epsilon)$. Approximating $\gamma_0$ within the linear-response regime permits further analytical insights. As explained in detail in Appendix~\ref{app:B},  linear response at zero temperature corresponds to weak energy dependence of the transmission function over the bias window, permitting the Sommerfeld expansion $\mathcal T(\epsilon) \approx \mathcal T_F + (\epsilon-\epsilon_F) \partial_\epsilon \mathcal T(\epsilon)\Big\vert_{\epsilon = \epsilon_F}$, where $\mathcal T_F \equiv \mathcal T(\epsilon_F)$ is the transmission function evaluated at the Fermi energy. With this expansion, the current and noise are given by $\expval{I}_{\text{LR}}^{T\to0} = (2e^2V/h)\mathcal T_F$ and $\expval{\expval{I^2}}_{\text{LR}}^{T\to0} = (4e^3V/h)\mathcal T_F(1-\mathcal T_F)$, respectively. Therefore, the following expression for the precision rate is obtained
\begin{equation}
\label{eq:gamma_zero}
    \gamma_{0} = \frac{ eV \mathcal T_F\left(\partial_\theta \ln\mathcal T_F\right)^2}{h(1-\mathcal T_F)}  = \frac{1}{2}\frac{\mathcal T_F}{1-\mathcal T_F} \left(\partial_\theta \ln\mathcal T_F\right)^2 \frac{G_0 V}{e},
\end{equation}
where $G_0 = 2e^2/h$ is the quantum of conductance. 
This expression has an elegant interpretation: each electron transmitted across the junction can be considered as a successful Bernoulli trial, with $\mathcal{T}_F$ the success probability. The first factor $\mathcal{T}_F/(1-\mathcal{T}_F)$ is then the signal-to-noise ratio (the mean squared divided by the variance) of the Bernoulli random variable. Under our assumption that $\mathcal{T}_F$ is approximately energy-independent,  the conductance is $G=G_0 \mathcal{T}_F$, so that the second factor is the relative sensitivity as seen also in Eq. \eqref{eq:preclin}. Finally, $G_0V/e$ represents  the frequency with which the Bernoulli trials are attempted.

\textit{Lorentzian transmission function.---}To illustrate the general formulas derived above, we first consider the specific example of a two-terminal quantum conductor with a Lorentzian transmission function, 
\begin{align}
   \mathcal T^\text{lor}\left( \epsilon\right) =  \frac{\Gamma^2}{\Gamma^2 + (\theta - \epsilon)^2}.
\end{align}This corresponds to a single-level quantum dot with energy $\theta$. The coupling $\Gamma$ characterises the width, with $2\Gamma$ the full width at half maximum. 

In Fig.~\ref{fig:linrep} (a), we show the precision rate as a function of $\theta$ computed with both the exact Landauer-B\"uttiker expressions (solid curves) and the linear-response result (dashed curves). Naturally, the linear-response approximation breaks down for $k_BT \lesssim eV$, so it is only displayed for the high-temperature regime shown in the bottom panel. Due to the form of the transmission function, the curves are symmetric around a local minimum, showing two equal peaks. Insets in the bottom panel of Fig.~\ref{fig:linrep} (a) show the conductance (left) and the relative sensitivity (right) in linear response. At larger coupling, the increase in conductance compensates for the lowering of the relative sensitivity, leading to higher precision. Beyond linear response, at low temperature (top panel of Fig.~\ref{fig:linrep} (a)), the precision rate is sharply peaked near $\theta = \mu_{L,R}$ and its maximum value is up to 4 orders of magnitude greater than in the linear-response regime. 

The zero-temperature limit \eqref{eq:gamma_zero} with a Lorentzian transmission function takes the following simple form,
\begin{align}\label{eq:lorT}
    \gamma_{0} = 2\frac{G_0V}{e}\left( \frac{\mathcal T^\text{lor}_F}{\Gamma}\right)^2,
\end{align}The maximum $\gamma^{\text{max}}_{0} = 2G_0V/(\Gamma^2e)$ lies at $\theta=\epsilon_F$. 
It can be checked that the above corresponds to the expansion of Eq. \eqref{eq:zerT} with a Lorentzian transmission function to lowest order in $eV/\Gamma$. Therefore, the approximation of smooth energy dependence for a Lorentzian transmission function implicitly corresponds to the strong coupling regime $eV\ll\Gamma$ between the conductor (the dot) and the reservoirs.

\textit{Multiple Lorentzians and the boxcar transmission function.---}We now consider a boxcar transmission function,
\begin{align}\label{eq:box}
    \mathcal T^{\text{box}}\left( \epsilon\right) = \Theta\left(\epsilon+\delta - \theta \right) - \Theta\left(\epsilon-\delta-\theta \right),
\end{align}where $2\delta$ is the width of $\mathcal T^{\text{box}}$ and $\theta$ represents an overall energy translation in the positive direction. 
As shown in Appendix \ref{app:B}, 
this transmission function can be seen as arising as a sum of sharply peaked Lorentzians under a certain limit. This allows us to build a smooth approximation for the boxcar by placing $N+1$ Lorentzians with width $\Gamma$, centred at evenly spaced points in the energy window $[-\delta,\delta]$, and appropriately normalising the sum. This corresponds to the transmission function of $N_d = N+1$ quantum dots,
\begin{align}
\label{boxcar}
    \mathcal T_N(\epsilon) & = \frac{1}{\mathcal N}\sum_{i=0}^{N_d-1}\frac{\Gamma^2}{\Gamma^2+(\epsilon_i-\epsilon)},
\end{align}
where $\epsilon_i = -\delta + i\Delta $ is the median of Lorentzian $i$, $\Delta=2\delta/(N_d-1)$
 and $\mathcal N = \sum_{i=0}^{N_d-1} \Gamma^2/(\Gamma^2 + \epsilon_i^2)$. While this function does not produce a boxcar exactly, in the regimes $\Gamma\ll\delta$ and $N_d\gg\delta/\Gamma$, it provides a reasonable approximation; see the inset of Fig.~\ref{fig:boxc} for an illustration and Appendix \ref{app:B} for details.

\begin{figure}
    \centering
    \includegraphics[width=\columnwidth]{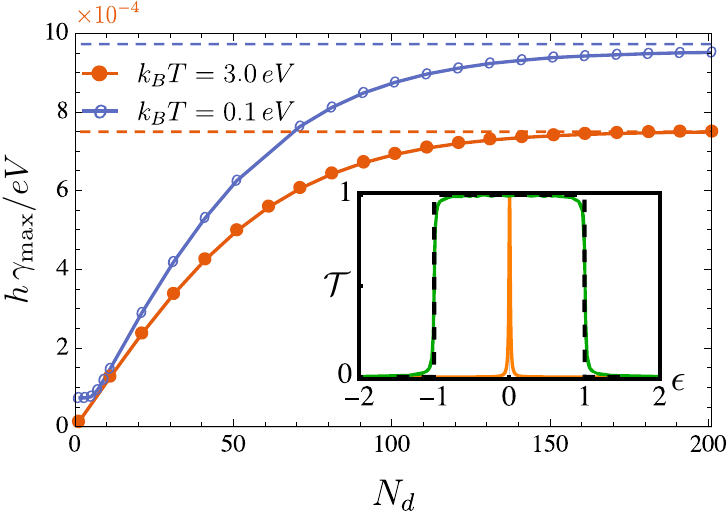}
    \caption{The maximum precision $\gamma_{\text{max}}$ (optimized over $\theta$) obtained with the full Landauer-Büttiker expressions, as a function of the number of Lorenztians $N$ added within the energy window $[-\delta,\delta]$. The dashed lines are obtained with a boxcar transmission function over the same energy window. Parameters: $k_BT=3\,eV$, $\delta=100\,\Gamma$. The inset shows $\mathcal T^{\text{box}}$ (dashed, black), a $\mathcal T^{\text{lor}}$ (orange) and $\mathcal T_N$ with $N_d=201$ (green).}
    \label{fig:Nlor}
\end{figure}

We now compare the precision obtained in the case of a single Lorentzian transmission function with that obtained with a sum of many such functions and the boxcar function of Eq.~\eqref{boxcar}. In Fig.~\ref{fig:linrep} (b) and (c), we show the precision rate as a function of $\theta$ in the case of multiple (specifically, a sum of 21) Lorentzians and a boxcar transmission function, within the same energy window $[-\delta,\delta]$, respectively. Note that we have appropriately normalised this energy such that $\delta=100\Gamma$ is held constant. This is necessary to ensure a smooth approximation to the boxcar function in the limit of large $N_d$, and to make a fair comparison with the boxcar case. While the general behaviours are often similar in the three cases, the precision rate in the case of $\mathcal T_N$ can show more than two peaks. In all coupling and temperature regimes, we find an increase in the precision rate for $\mathcal T_N$, and even further for $\mathcal T^{\text{box}}$, as compared to the single Lorentzian case. 

To assess how precision depends on the shape of the transmission function, we vary $N_d$ to smoothly interpolate between a sharply peaked (Lorentzian) and a flat (boxcar) transmission function. For each value of $N_d$, we compute the optimal precision rate (maximized over $\theta$) and plot the results in Fig. \ref{fig:Nlor} for two different temperatures. These temperatures are chosen such that the red curve corresponds to the linear-response regime ($eV\ll k_BT$), whereas the blue curve is far away from it. We again choose $\Gamma\ll\delta$ to simulate a boxcar as precisely as possible at large $N_d$. This is necessary to ensure that the solid curves obtained with $\mathcal T_N$ almost match with the dashed curves obtained with the boxcar. At both temperatures considered in Fig.~\ref{fig:Nlor}, the precision rate increases monotonically with $N_d$, saturating to the maximum value as $\mathcal T_N$ approaches a boxcar function at large $N_d$. Interestingly, moreover, we show in Appendix~\ref{app:E} that the precision rate for the boxcar function diverges at zero temperature, originating from a diverging signal-to-noise ratio in this situation. From this investigation, we conclude that the boxcar transmission function is optimal for current-based parameter estimation in two-terminal mesoscopic conductors within and beyond the linear-response regime at any temperature.

\textit{Conclusion.---}In this work, we have formulated precision bounds within a Landauer-B\"uttiker formalism, allowing to assess metrology limits in two-terminal quantum coherent conductors. Analytical expressions are derived in two limits, at thermal equilibrium (Eq.~\eqref{eq:preclin}) and at zero temperature (Eq.~\eqref{eq:gamma_zero}). These results enable a transparent understanding of the key parameters and energy scales for optimization of the precision rate. In addition, we compare current-based metrology between different transmission functions, interpolating between a single-peak Lorentzian and a boxcar function. Our results indicate that, in the linear response regime, strong coupling enhances precision by boosting sensitivity, while outside of linear response we find that intermediate coupling strengths $\Gamma\sim k_BT$ are optimal. Meanwhile, low temperature is generally beneficial because it suppresses current noise. Furthermore, we numerically find that the boxcar transmission function is optimal for parameter estimation in quantum coherent mesoscopic conductors, in all parameter regimes considered. Interestingly, boxcar transmission functions have been found to be optimal for the efficiency~\cite{Whitney2014, *Whitney2015,Tesser2023} and constancy~\cite{Timpanaro2025} of thermoelectric power generation in similar setups. Such an energy-independent transmission probability can be approximated by a linear chain of quantum dots with carefully engineered couplings~\cite{Ehrlich2021}.

While our exact analysis within the Landauer-B\"uttiker formalism enabled optimization of precision across arbitrary regimes of coupling strength and nonequilibrium bias, this approach neglects interactions between charge carriers. Our results therefore complement those of Ref.~\cite{Mihailescu2024}, which derived the quantum Fisher information for interacting systems within adiabatic linear-response theory, thus assuming weak deviation from equilibrium. We also emphasize that our precision analysis is restricted to a specific, accessible observable: the charge current. This is readily measured in experiments but it may not be an optimal observable in the metrological sense of saturating the quantum Cram\'er-Rao bound~\cite{Paris2009}. An interesting future research problem is thus to find the optimal precision in far-from-equilibrium settings, as dictated by the quantum Fisher information of the entire system-reservoir setup, and establish whether it can be achieved by realistic measurements. Since currents are defined by measurements at (at least) two times, this calls for an inherently multi-time approach beyond the Markov approximation, e.g.~as in Refs.~\cite{Tsang2011,Tsang2013,Blasi2025}. This may shed light on whether non-equilibrium steady-state entanglement~\cite{Khandelwal2020,Heineken2021,Diotallevi2024,Mortimer2024,Khandelwal2025} or quantum violations of classical precision bounds~\cite{ptaszynski_coherence-enhanced_2018,brandner_thermodynamic_2018,guarnieri_thermodynamics_2019, hasegawa_quantum_2020, van_vu_thermodynamics_2022, prech_entanglement_2023, Moreira2025, Brandner2025, Blasi2025a, Palmqvist2025} can provide further advantages for current-based metrology.

\textit{Acknowledgments.---} We thank Gianmichele Blasi, George Mihailescu, Andrew K. Mitchell, and Patrick P. Potts for insightful discussions. S.K. acknowledges support from the Swiss National Science Foundation Grant No.
P500PT\textunderscore222265 and the Knut and Alice Wallenberg Foundation through the Wallenberg Center for Quantum Technology (WACQT). G. H. acknowledges financial support from the NCCR SwissMAP from the Swiss National Science Foundation. M.T.M. is supported by a Royal Society University Research Fellowship. This project is co-funded by the European Union (Quantum Flagship project ASPECTS, Grant Agreement No.~101080167) and UK Research \& Innovation (UKRI). Views and opinions expressed are however those of the authors only and do not necessarily reflect those of the European Union, Research Executive Agency or UKRI. Neither the European Union nor UKRI can be held responsible for them.
GTL is supported by the U.S. Department of Energy (DOE), Office of Science, Basic Energy Sciences (BES) under Award No. DE-SC0025516.

\bibliography{references}

\clearpage
\newpage
\appendix

\onecolumngrid

\section{Charge and current statistics}
\label{app:charge_statistics}

We consider a two-point measurement of the particle number $\hat{N}_R$ of the right reservoir (equivalent results are obtained for the left reservoirs). Formally, this is described by two projective measurements at time $t=0$ and $t=\tau$, leading to outcomes $n$ and $m$ with corresponding joint probability $P(m,n) = {\rm tr}[\hat{\Pi}_m \hat{U}(\tau) \hat{\Pi}_n \hat{\rho} \hat{\Pi}_n \hat{U}^\dagger(\tau)]$, where $\hat{\Pi}_n$ is the projector onto the charge state with precisely $n$ electrons in the reservoir, $\hat{U}(\tau)$ is the unitary time evolution operator over the time interval $\tau$, and $\hat{\rho}$ is the initial state. The probability that an amount of charge $Q =e q$ is transferred, with $q$ an integer, is then
\begin{equation}
\label{P_of_Q}
    P(Q=eq) = \sum_{m,n} {\rm tr}[\hat{\Pi}_m \hat{U}(\tau) \hat{\Pi}_n \hat{\rho} \hat{\Pi}_n \hat{U}^\dagger(\tau)]\delta_{q,m-n}.
\end{equation}
To obtain the moments of $Q$, it is convenient to express this in terms of the characteristic function~\cite{esposito_nonequilibrium_2009}
\begin{equation}
    \label{chi_Q}
    \chi(\lambda) = \sum_{q=-\infty}^\infty P(Q=eq) e^{i\lambda q} = {\rm tr}[\hat{U}^\dagger(\tau)e^{i\lambda \hat{N}_R} \hat{U}(\tau) e^{-i\lambda \hat{N}_R} \hat{\rho}'],
\end{equation}
where $\hat{\rho}' = \sum_n \hat{\Pi}_n \hat{\rho}\hat{\Pi}_n$. 
The moments of the transferred charge now follow as 
\begin{equation}
    \label{Q_moments}
    \langle Q^k\rangle = (-ie)^k \left.\frac{\partial^k \chi}{\partial \lambda^k}\right|_{\lambda=0}.
\end{equation}

The first two moments are found to be 
\begin{align}
    \label{first_moment}
    &\langle Q\rangle =  \left\langle \hat{N}_R(\tau)\right\rangle - \left\langle \hat{N}_R(0)\right\rangle, \\
    \label{second_moment}
    & \langle Q^2\rangle = \left\langle \left(\hat{N}_R(\tau)- \hat{N}_R(0)\right)^2\right\rangle,
\end{align}
where $N_R(\tau) = \hat{U}^\dagger(\tau)\hat{N}_R\hat{U}(\tau)$ is the charge operator in the Heisenberg picture, and we denote quantum expectation values on the right-hand side by $\langle \bullet\rangle = {\rm tr}[\bullet\hat{\rho}']$. Equation~\eqref{second_moment} is obtained by exploiting the property $[\hat{N}_R,\hat{\rho}']=0$. To connect these expressions to the charge current, we define the current operator flowing into the right reservoir,
\begin{equation}
    \label{current_operator}
    \hat{I}_R(t) = \frac{d}{dt} \hat{N}_R(t).
\end{equation}
Then, the first moment can be expressed as
\begin{equation}
    \langle Q\rangle = \int_0^\tau dt\, \left\langle \hat{I}_R(t)\right\rangle \approx \tau \langle I\rangle.
\end{equation}
The final approximation holds for $\tau \gg t_{\rm rel}$, where $t_{\rm rel}$ is the relaxation time after which the current becomes stationary, $\langle \hat{I}_R(t)\rangle \to \langle I\rangle.$ The  variance can similarly be expressed as
\begin{equation}
    \label{charge_variance}
    {\rm Var}[Q] = \int_0^\tau dt \int_0^\tau dt' \left [\left\langle \hat{I}_R(t) \hat{I}_R(t') \right\rangle - \left\langle \hat{I}_R(t) \right\rangle \left\langle\hat{I}_R(t') \right\rangle \right].
\end{equation}
For $\tau\gg t_{\rm rel}$, the above correlation function becomes stationary, i.e.~dependent only on the time difference $u=t-t'$, and we can write
\begin{equation}
    \label{charge_variance_stationary}
    {\rm Var}[Q] \approx \int_{-\tau}^0 dt\, (\tau+t) C_R(t) + \int_{0}^\tau dt\, (\tau-t)  C_R(t),
\end{equation}
where $C_R(t) = \langle \hat{I}_R(t) \hat{I}_R(0) \rangle - \left\langle I\right \rangle^2 $ is the stationary two-point correlation function of the current. Assuming that $\tau \gg t_{\rm cor}$, where $t_{\rm cor}$ is the time over which the correlation function decays to zero, we can further approximate
\begin{equation}
    \label{charge_variance_correlation_time}
    {\rm Var}[Q] \approx \tau \int_{-\infty}^\infty dt\, C_R(t) \equiv \tau \langle \langle I^2\rangle \rangle.
\end{equation}
This defines the DC current noise $\langle \langle I^2 \rangle \rangle = S_R(0)$, where 
\begin{equation}
    \label{current_noise_spectrum}
    S_R(\omega) = \int_{-\infty}^\infty dt\, e^{i\omega t} C_R(t)
\end{equation}
is the current noise spectrum. Note that in the steady state, the mean and DC noise of the current are the same in both reservoirs~\cite{Buttiker1992}, and therefore we drop the subscript $R$ in the main text.

\section{Fisher information}
\label{app:Fisher}

In general, the estimation precision of current-based metrology is fundamentally constrained by the Cram\'er-Rao bound~\cite{kay_fundamentals_2013}
\begin{equation}\label{CRB}
    {\rm Var}[\check{\theta}] \geq \frac{1}{F(\theta)},
\end{equation}
where $F(\theta)$ is the Fisher information of the charge distribution:
\begin{equation}
    \label{Fisher_info}
    F(\theta) = \int dQ \, P(Q) \left[\partial_\theta \ln P(Q)\right]^2.
\end{equation}
Therefore, the precision of the estimator discussed in the main text above Eq.~\eqref{eq:prec}, provides a lower bound on the Fisher information via Eq.~\eqref{CRB}, i.e.~$F(\theta) \geq \gamma\tau $, where $\gamma$ is the precision rate defined in Eq.~\eqref{eq:prec}.

This bound is saturated at long times if the measured charge distribution is approximately Gaussian, so that we can write
\begin{equation}
\label{Gaussian_approx}
P(Q) \approx \sqrt{2\pi \tau \langle \langle I^2\rangle\rangle} \exp\left[ -\frac{(Q - \tau\langle I\rangle)^2}{2 \tau\langle\langle I^2\rangle\rangle }\right].
\end{equation}
This is expected to be a good approximation if the charge distribution is unimodal and if the integration time $\tau$ is much longer than the decay time of all current auto-correlations (including correlation functions above second order). The Fisher information for the distribution~\eqref{Gaussian_approx} reads as
\begin{equation}
    \label{Fisher_Gaussian}
    F(\theta) = \frac{\tau(\partial_\theta \langle I\rangle)^2}{ \langle \langle I^2\rangle\rangle} + \left(\tfrac{1}{2}\partial_\theta \ln \langle\langle I^2\rangle\rangle\right)^2.
\end{equation}
For large $\tau$, the second term is subleading and we recover the precision rate in Eq.~\eqref{eq:prec}.

\section{Landauer-Büttiker expressions}
\label{app:B}
Under the Landauer-Büttiker formalism, the average steady-state current is given by \cite{Blanter2000},
\begin{align}\label{eq:J}
    \expval{I} = 2\frac{e}{h}\int d\epsilon\, \mathcal T\left(\epsilon\right) \left[f_R\left(\epsilon\right) -f_L\left(\epsilon\right) \right],
\end{align}
and the steady-state noise or the zero-frequency component of the power spectrum,
\begin{align}\label{eq:Jf}
\expval{\expval{I^2}} = 4\frac{e^2}{h}\int d\epsilon\,\left[\mathcal{T}\left(\epsilon\right) \left(1-\mathcal T\left( \epsilon\right) \right) \mathcal F_1\left(\epsilon\right) + \mathcal T\left( \epsilon\right) \mathcal F_2\left(\epsilon\right)\right],
\end{align}where 
\begin{equation}
\begin{aligned}
  &\mathcal F_1\left(\epsilon \right)\coloneqq \left(f_R\left(\epsilon \right) - f_L\left(\epsilon \right) \right)^2 \quad \text{and}\quad  &\mathcal F_2\left(\epsilon \right) \coloneqq f_L\left(\epsilon \right)\left(1-f_L\left(\epsilon \right) \right) + f_R\left(\epsilon \right)\left(1-f_R\left(\epsilon \right) \right),
  \end{aligned}
\end{equation}with the Fermi-Dirac distribution,
$f_{L/R}(\epsilon) \equiv f\left(\epsilon,\mu_{L/R},T_{L/R} \right) = 1/{e^{\left(\epsilon - \mu_{L/R} \right)/k_BT_{L/R}}+1},$. Throughout our work, we have chosen $\mu_R =\mu=\epsilon_F+eV$, $\mu_L=\epsilon_F$ and $T_L=T_R=T$.

\subsection{Linear response}
The goal in this section is to derive linear-response expressions for the current, noise and precision that are valid for both regimes considered in this work, i.e., $k_BT\gg eV$ and $T\to0$. 
\subsubsection{Current}

A general linear-response expression for average current can naturally be derived by expanding it in the chemical potential around the Fermi energy $\epsilon_F$,
\begin{align}
    \expval{I} = \sum_{n=0}^\infty \frac{(eV)^n}{n!}\partial^n_\mu\expval{I}\rvert_{\mu=\epsilon_F} . 
\end{align}While this works directly for $k_BT\gg eV$, it is important to note that the derivative of the Fermi distribution can be singular for $T\to0$. A natural way to evade this problem is to first note that the derivative of the Fermi distribution satisfies $\partial^n f/\partial \mu^n =(-1)^n \partial^n f/\partial \epsilon^n$. Then, the derivative can be transferred from the Fermi distribution to the transmission function through integration by parts. Since we want a linear-response expression, we focus only on the first two terms in the above series. The zeroth-order term is naturally zero, while the first-order term takes the form
\begin{align}
    2\frac{e^2V}{h}\int d\epsilon \mathcal T(\epsilon) \partial_\mu f(\epsilon,\mu,T)\lvert_{\mu=\epsilon_F} = -2\frac{e^2V}{h}\int d\epsilon \partial_\epsilon\left(\mathcal T(\epsilon)  f(\epsilon,\epsilon_F,T)\right) +2\frac{e^2V}{h}\int d\epsilon \partial_\epsilon\mathcal T(\epsilon)  f(\epsilon,\epsilon_F,T)
\end{align}The cases considered in this work involve transmission functions that vanish at $\pm\infty$. Therefore the first term vanishes and we are left with 
\begin{align}\expval{I}_{\text{LR}} = 
2\frac{e^2V}{h}\int d\epsilon \partial_\epsilon\mathcal T(\epsilon)  f(\epsilon,\epsilon_F,T)\label{eq:curapp}
\end{align}The above expression is valid at all temperatures. However, the consistency of this approach relies on the derivatives of all orders of the transmission function being smooth. Within linear response, this means that the dependence of the transmission function on energy must be weak within the bias window. This can be seen in the following manner.
\begin{align}
    \expval{I} = 2\frac{e}{h}\int d\epsilon \mathcal T(\epsilon) \left(f(\epsilon,\epsilon_F+eV,T) - f(\epsilon,\epsilon_F,T) \right) = 2\frac{e}{h}\int d\epsilon \mathcal T(\epsilon)f(\epsilon-eV,\epsilon_F,T) - 2\frac{e}{h}\int d\epsilon \mathcal T(\epsilon) f(\epsilon,\epsilon_F,T),
\end{align}where we have moved the voltage bias in the energy argument in the Fermi distribution. We now make the variable change $\epsilon-eV\to\epsilon$ in the first integral to obtain
\begin{align}
    \expval{I} = 2\frac{e}{h}\int d\epsilon \mathcal T(\epsilon+eV)f(\epsilon,\epsilon_F,T) - 2\frac{e}{h}\int d\epsilon \mathcal T(\epsilon) f(\epsilon,\epsilon_F,T).
\end{align}To obtain the above up to linear order in $eV$, we expand the transmission function to the first order, giving us
\begin{align}
    \expval{I}_{\text{LR}} = 2\frac{e}{h}\int d\epsilon \left(\mathcal T(\epsilon) +eV\mathcal \partial_{\epsilon}\mathcal T(\epsilon)\right)f(\epsilon,\epsilon_F,T) - 2\frac{e}{h}\int d\epsilon \mathcal T(\epsilon) f(\epsilon,\epsilon_F,T) = 2\frac{e^2V}{h} \int d\epsilon \partial_\epsilon\mathcal T(\epsilon) f(\epsilon,\epsilon_F,T),
\end{align}which is identical to Eq. \eqref{eq:curapp}.

Specifically for $k_BT\gg eV$ and $T\to0$, we obtain the expressions used in the main text,
\begin{equation}
\label{eq:x1}
 k_BT\gg eV:\quad    \langle I\rangle_{\text{LR}} = GV, \quad G\coloneqq -2\frac{e^2}{h} \int d\epsilon  \mathcal T(\epsilon) \partial_\epsilon f(\epsilon,\epsilon_F,T) = 2\frac{e^2}{h} \int d\epsilon \partial_\epsilon \mathcal T(\epsilon) f(\epsilon,\epsilon_F,T)  
 \end{equation}
 \begin{equation}\label{eq:x2}
 T\to 0 : \quad \langle I\rangle_{\text{LR}}^{T\to0} = 2\frac{e^2 V}{h}\int_{-\infty}^{\epsilon_F} d\epsilon\partial_\epsilon\mathcal T(\epsilon) = 2\frac{e^2V}{h}\mathcal T_F = G_0V \mathcal T_F
 \end{equation}where $G_0=2e^2/h$ is the conductance quantum and $\mathcal T_F \equiv \mathcal T(\epsilon_F)$, i.e., the transmission function evaluated at the Fermi energy. We have utilised the property that at zero temperature, the Fermi distributions behave as Heaviside functions, $f_\alpha(\epsilon) = \Theta(\mu_\alpha - \epsilon)$.
Alternatively, the conductance $G$ can be expressed as 
\begin{align}\label{eq:condalt}
     G = 2\frac{e^2}{h} \int d\epsilon \partial_\epsilon \mathcal T(\epsilon) f(\epsilon,\epsilon_F,T) = - 2\frac{e^2}{h} \int d\epsilon \mathcal T(\epsilon) \partial_\epsilon f(\epsilon,\epsilon_F,T) = \frac{e^2}{h}\int d\epsilon \mathcal{T}(\epsilon) \frac{1}{k_BT\left(1+\cosh\left(\frac{\epsilon-\epsilon_F}{k_BT} \right)\right)},
 \end{align}where we have again utilised $\mathcal T(\pm\infty) = 0$.  
\subsubsection{Noise}
Next, we consider the noise in linear response with an expansion of the form,
\begin{align}
    \expval{\expval{I^2}} = \sum_{n=0}^\infty \frac{(eV)^n}{n!}\partial^n_\mu\expval{\expval{I^2}}\rvert_{\mu=\epsilon_F} . 
\end{align}We follow a similar procedure as presented above. First, consider the zeroth order term.
\begin{align}
    \expval{\expval{I^2}}\lvert_{\mu = \epsilon_F} = 4\frac{e^2}{h} \left[\int d \epsilon \mathcal{ T}(\epsilon)(1-\mathcal T(\epsilon))\mathcal F_1\right] \Bigg{\rvert}_{\mu = \epsilon_F} + 4\frac{e^2}{h} \left[ \int d\epsilon \mathcal T(\epsilon) \mathcal F_2\right]\Bigg\rvert_{\mu = \epsilon_F}
\end{align}Note that the first term simply evaluates to zero at $\mu=\epsilon_F$, while the second term is non-zero. We can therefore write
\begin{equation}
\begin{aligned}
    \expval{\expval{I^2}}\lvert_{\mu = \epsilon_F} &=  \frac{4e^2}{h}  \int d\epsilon \mathcal T(\epsilon)2 f_L\left( 1-f_L\right) \\
    & =  \frac{4e^2}{h}  \int d\epsilon \mathcal T(\epsilon) \frac{1}{\left(1+\cosh\left(\frac{\epsilon-\epsilon_F}{k_B T} \right)\right)} \\
    & =  4k_B T G,
\end{aligned}
\end{equation}where in the last step we have utilised the expression \eqref{eq:condalt} for the conductance, obtaining the expression for thermal noise in linear response, which is zeroth-order in $eV$. This term is zero at $T=0$ and non-zero at all $T>0$. Since we have a finite zeroth-order contribution to the noise for $T>0$, we do not require a higher (first) order correction in this regime, as it will be negligible.

Now consider the first-order correction. As stated above, we are only looking for this correction for the case $T=0$. For convenience, we set $\mathcal T(\epsilon)(1-T(\epsilon))\equiv \tilde{\mathcal{T}}(\epsilon)$. We then have the first-order correction,
\begin{align}
    eV \partial_\mu\expval{\expval{I^2}}\lvert_{\mu=\epsilon_F} =  \frac{4e^3 V}{h} \left[\partial_\mu\int d \epsilon \tilde{\mathcal  T}(\epsilon)(f_R-f_L)^2\right] \Bigg{\rvert}_{\mu = \epsilon_F} +\frac{4e^3 V}{h} \left[\partial_\mu\int d \epsilon \mathcal{ T}(\epsilon)(f_R(1-f_R)+f_L(1-f_L))\right] \Bigg{\rvert}_{\mu = \epsilon_F}
\end{align}First consider the left term. 
For $T\to0$, the the difference of the Fermi functions behaves as a square function in the bias window. Assuming that the transmission function is continuous, we have
\begin{equation}
\begin{aligned}
    4\frac{e^3 V}{h} \left[\partial_\mu\int d \epsilon \tilde{\mathcal  T}(\epsilon)(f_R-f_L)^2\right] \Bigg{\rvert}_{\mu = \epsilon_F}  &\stackrel{T\to0}{=}4\frac{e^3 V}{h} \left[\partial_\mu\int_{\epsilon_F}^{\mu} d\epsilon \tilde{\mathcal T}(\epsilon)\right]\Bigg\rvert_{\mu=\epsilon_F} = 4\frac{e^3V}{h}\mathcal T_F(1-\mathcal T_F)
\end{aligned}
\end{equation}Now consider the second term, 
\begin{equation}
    \begin{aligned}
        4\frac{e^3 V}{h} \left[\partial_\mu\int d \epsilon \mathcal{ T}(\epsilon)(f_R(1-f_R)+f_L(1-f_L))\right] \Bigg{\rvert}_{\mu = \epsilon_F} \stackrel{T\to0}{=} 0
    \end{aligned}
\end{equation} We finally have 
\begin{align}
    \expval{\expval{I^2}}_{\text{LR}} = 4\frac{e^3 V}{h} \left[\partial_\mu\int d \epsilon \tilde{\mathcal  T}(\epsilon)(f_R-f_L)^2\right] \Bigg{\rvert}_{\mu = \epsilon_F} + 4k_B T G. 
\end{align}  
The above equation can be used to obtain linear-response expressions for the noise in the two relevant limits,
\begin{align}\label{eq:y1}
k_BT\gg eV:\quad \expval{\expval{I^2}}_{\text{LR}} = 4k_BT G 
\end{align}
\begin{align}\label{eq:y2}
    T\to0: \quad \expval{\expval{I^2}}_{\text{LR}}^{T\to0} = 4\frac{e^3 V}{h} \left[\partial_\mu\int d \epsilon \tilde{\mathcal  T}(\epsilon)(f_R-f_L)^2\right] \Bigg{\rvert}_{\mu = \epsilon_F} = 4\frac{e^3V}{h}\mathcal T_F(1-\mathcal T_F),
\end{align}where in the second line, we have utilised the zero-temperature property of the Fermi function and then have evaluated the integral, keeping terms up to linear order in $eV$. 

Next we consider linear-response expressions for the precision. Eq. (2) in the main text can be directly obtained from Eqs. \eqref{eq:x1} and \eqref{eq:y1}. The procedure to obtain Eqs. (3) and (4) is given in detail below.

\subsection{Zero-temperature limit of the precision rate}
As mentioned previously, at zero temperature, the Fermi distributions behave as Heaviside functions, $f_\alpha(\epsilon) = \Theta(\mu_\alpha - \epsilon)$. This sets a finite energy window of width $eV$ through which charge transport takes place. The current and shot noise become,
\begin{align}
    \langle I \rangle_{T\to0} = 2\frac{e}{h}\int_{\epsilon_F}^{\epsilon_F + eV} d\epsilon \, \mathcal T(\epsilon)\,,\quad
    \langle\langle I^2 \rangle\rangle_{T\to0} =4\frac{e^2}{h}\int_{\epsilon_F}^{\epsilon_F + eV} d\epsilon\, \mathcal T(\epsilon)  (1-\mathcal T(\epsilon)). 
\end{align}
Note that these expressions are valid beyond linear response. The zero-temperature precision takes the form,
\begin{eqnarray}\label{eq:zerT1}
    \gamma_{0} = \frac{1}{h} \frac{\left(\int_{\epsilon_F}^{\epsilon_F + eV}  d\epsilon \partial_\theta \mathcal T(\epsilon)\right)^2}{\int_{\epsilon_F}^{\epsilon_F + eV}  d\epsilon \mathcal T(\epsilon)(1-\mathcal T(\epsilon))}\,,
\end{eqnarray}where we have assumed that the $\theta$-dependence lies in the transmission function. To gain further analytical insights, we make a linear-response approximation. As previously mentioned, this corresponds to taking a smooth energy-dependence of the transmission function through the expansion, $  \mathcal T(\epsilon) \sim \mathcal T_F + (\epsilon-\epsilon_F) \, \partial_\epsilon\mathcal T(\epsilon)\vert_{\epsilon = \epsilon_F} \equiv \mathcal T_F + (\epsilon-\epsilon_F) \, \partial_\epsilon \mathcal T_F$.  
The current can therefore be simplified as 
\begin{align}\label{eq:cur0app}
    \expval{I}_{\text{LR}}^{T\to0} = 2\frac{e}{h}\int_{\epsilon_F}^{\epsilon_F + eV} d\epsilon \, \mathcal T(\epsilon) \approx 2\frac{e}{h}\int_{\epsilon_F}^{\epsilon_F + eV} d\epsilon \mathcal T_F + (\epsilon-\epsilon_F) \, \partial_\epsilon \mathcal T_F = 2\frac{e^2V}{h}\mathcal T_F,
\end{align}where we have disregarded higher-order terms in $eV$. This matches the previously obtained expression \eqref{eq:x2}, which was obtained by taking the linear-response limit before the zero-temperature one. The noise is obtained in the precisely the same way as for Eq. \eqref{eq:y2},
\begin{align}\label{eq:noi0app}
    \expval{\expval{I^2}}^{T\to0}_{\text{LR}} = \frac{4e^3V}{h}\mathcal T_F(1-\mathcal T_F).
\end{align}

Using Eqs. \eqref{eq:cur0app} and \eqref{eq:noi0app}, the linear response expression for zero-temperature precision can expressed as
\begin{eqnarray}
\label{eq:gamma_zero1}
    \gamma^{\text{LR}}_{0} = \frac{\left(\partial_\theta\expval{I}_{\text{LR}}^{T\to0}\right)^2}{\expval{\expval{I^2}}_{\text{LR}}^{T\to0}} = 
\frac{eV}{h} \, \frac{\mathcal T_F}{1-\mathcal T_F} \left(\partial_\theta \ln\mathcal T_F\right)^2 =\frac{1}{2}\frac{\mathcal T_F}{1-\mathcal T_F}\left(\partial_\theta \ln\mathcal T_F\right)^2\frac{G_0V}{e}.
\end{eqnarray}

\section{Boxcar as a sum of Lorentzian functions}
\label{app:C}
\begin{figure}
    \includegraphics[width = 0.8\textwidth]{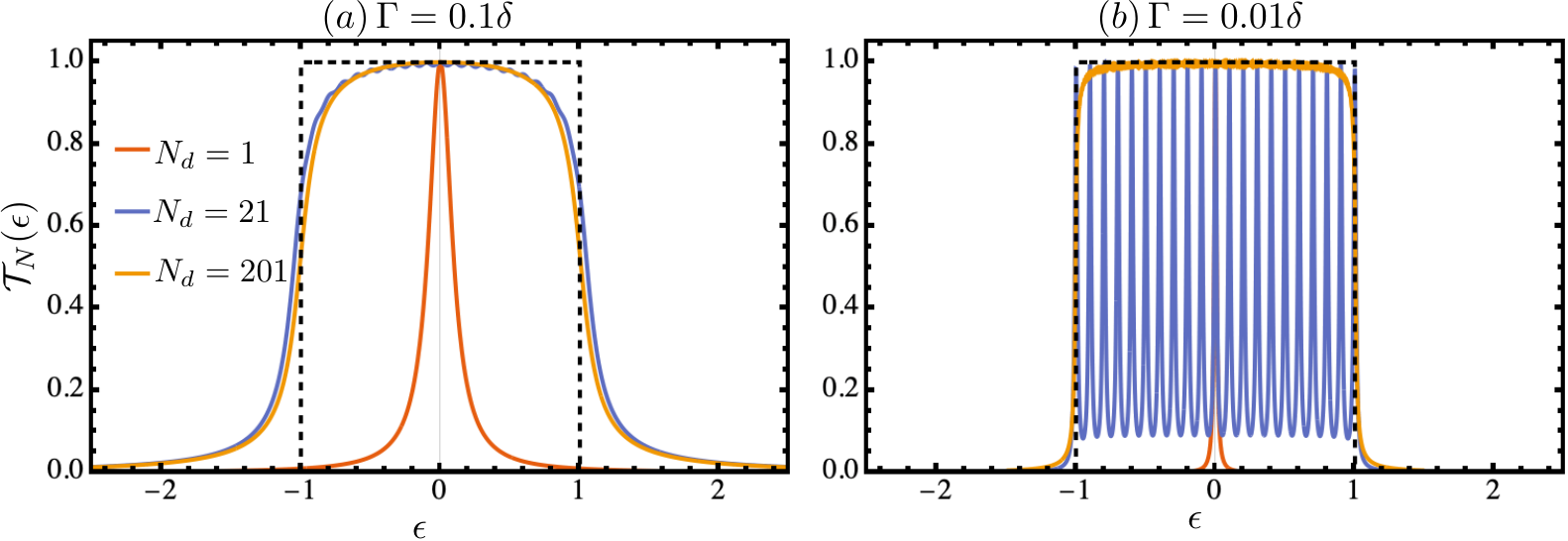}\caption{$\mathcal T_N$ as a function of $\epsilon$ for (a) $\Gamma=0.1\delta$ and (b) $\Gamma=0.01\delta$. Other parameters: $\delta=1$ and $\theta=0$. }\label{fig:boxc}
\end{figure}

In this appendix, we show how to construct a smooth approximation for a boxcar function, $\mathcal T^{\text{box}}(\epsilon)$, defined as 
\begin{align}
\mathcal T^{\text{box}}(\epsilon) = \begin{cases}
1, \quad -\delta<\epsilon<\delta\\
0,\quad \text{otherwise}
\end{cases}
\end{align}We can write
\begin{equation}
\begin{aligned}
\mathcal T^{\text{box}}(\epsilon) = \int d\epsilon^\prime\, \mathcal T^{\text{box}}(\epsilon^\prime)\delta(\epsilon-\epsilon^\prime)=\int_{-\delta}^{\delta} d\epsilon^\prime\,\delta(\epsilon-\epsilon^\prime) = \lim_{\Gamma\to0} \int_{-\delta}^{\delta}d\epsilon^\prime\,\frac{1}{\pi} \frac{\Gamma}{\Gamma^2  + (\epsilon^\prime-\epsilon)^2} 
\end{aligned}
\end{equation}Writing the last integral as a Riemann sum, the boxcar function can be seen as a sum of Lorentzian functions spread over the range of the boxcar function,
\begin{align}
   \mathcal T^{\text{box}}(\epsilon) = \lim_{\Gamma\to0}\lim_{N\to\infty} \sum_{i=0}^{N-1} \frac{\Delta}{\pi} \frac{\Gamma}{\Gamma^2 + \left(\epsilon_i - \epsilon\right)^2}, 
\end{align}with  $\Delta = 2\delta/N$ and  $\epsilon_i = -\delta + i\Delta $. Therefore, a boxcar function can be approximately simulated with a large number of sharply peaked Lorentzian functions. Note that the above sum is only normalized to (or peaked at) 1 for $N\to\infty$. In practice, to approximate a boxcar function for numerical calculations, we have taken a sum of uniformly spaced Lorentzian functions, each centered in the interval $[-\delta,\delta]$ and have appropriately normalized this sum such that the peak lies at 1.  Taking inspiration from the trapezoidal rule, consider the following sum,
\begin{align}
    \bar{\mathcal T}_N(\epsilon) = \frac{1}{\bar{\mathcal N}}\left(\sum_{i=1}^{N-1}\frac{\Gamma^2}{\Gamma^2+(\epsilon_i-\epsilon)^2}+\frac{1}{2}\left( \frac{\Gamma^2}{\Gamma^2+(\delta-\epsilon)^2}+\frac{\Gamma^2}{\Gamma^2+(\delta+\epsilon)^2}\right)\right),
\end{align}where $\epsilon_i$ is as defined above and $\tilde{\mathcal N} = \sum_{i=1}^{N} \Gamma^2/(\Gamma^2 + \epsilon_i^2)$ is a positive factor that normalizes the peak of the sum to 1. For numerical calculations in the main text, we have fixed $\Gamma\ll \delta$. Then, for large $N$, this sum approximates a boxcar function. The correspondence between $\mathcal T^\text{box}$ and $\bar{\mathcal T}_N$ also be seen by evaluating $\bar{\mathcal N}$ for large $N$,
\begin{align}
    \bar{\mathcal N} \Delta = \Delta\sum_{i=1}^N \frac{\Gamma^2}{\Gamma^2+\epsilon_i^2} = \Delta\sum_{i=1}^{N-1}\frac{\Gamma^2}{\Gamma^2+\epsilon_i^2} + \frac{\Delta}{2}\left( \frac{\Gamma^2}{\Gamma^2+(-\delta)^2} + \frac{\Gamma^2}{\Gamma^2+\delta^2}\right)\approx \int_{-\delta}^\delta d\epsilon\frac{\Gamma^2}{\Gamma^2+\epsilon^2} = 2\Gamma \tan^{-1}\left( \frac{\delta}{\Gamma}\right) \approx \Gamma\pi,
\end{align}where we have used the trapezoidal rule and the approximation $\tan^{-1}(\delta/\Gamma)\approx \pi/2$ for $\Gamma\ll\delta$. Therefore, for large $N$ and small $\Gamma$, we obtain $\bar{\mathcal N} \approx \Gamma\pi/\Delta$. Finally, $\bar{\mathcal T}_N$ can be expressed as
\begin{equation}
\begin{aligned}
    \bar{\mathcal T}_N(\epsilon) &\approx \left(\sum_{i=1}^{N-1}\frac{\Delta}{\Gamma\pi}\frac{\Gamma^2}{\Gamma^2+(\epsilon_i-\epsilon)}+\frac{1}{2}\frac{\Delta}{\Gamma\pi}\left( \frac{\Gamma^2}{\Gamma^2+(\delta-\epsilon)^2}+\frac{\Gamma^2}{\Gamma^2+(\delta+\epsilon)^2}\right)\right) \\&\approx \int_{-\delta}^{\delta}d\epsilon^\prime\,\frac{1}{\pi} \frac{\Gamma}{\Gamma^2  + (\epsilon^\prime-\epsilon)^2},
\end{aligned}
\end{equation}where in the second step, we have again used the trapezoidal approximation. The exact scaling of the error in this approximation, $\lvert \mathcal T^{\text{box}} - \bar{\mathcal T}_N\rvert$ can be estimated by means of the Euler-Maclaurin formula.
For the regime $\Gamma\ll\delta$ and $N\gg\delta/\Gamma$, a further simplification can be made. The end points of the summation $\bar{\mathcal T}_N$, constitute a contribution that is only $\sim\mathcal O(\frac{\Gamma^2}{N(\delta\pm\epsilon)^2})$, which is small away from the edges $\epsilon=\pm\delta$. In the immediate vicinity of the edges, the edge corrections are $\sim\mathcal O(\frac{\delta}{\Gamma N})$, which is small when $N\gg \delta/\Gamma$. Therefore, if $N\gg\delta/\Gamma$, the terms with the $1/2$ factor can be disregarded completely or the factor $1/2$ can be removed without introducing a significant error. We therefore approximate this sum as
\begin{align}
     \bar{\mathcal T}_N(\epsilon)\approx \mathcal T_N(\epsilon) = \frac{1}{\mathcal N}\sum_{i=0}^{N}\frac{\Gamma^2}{\Gamma^2+(\epsilon_i-\epsilon)},
\end{align}with $\mathcal N = \sum_{i=0}^N \Gamma^2/(\Gamma^2 + \epsilon_i^2)$ and $\epsilon_i$ as defined above. This corresponds to the transmission function of $N_d=N+1$ non-interacting quantum dots evenly spaced in energy. $\mathcal T_N(\epsilon)$ is depicted in Fig. \ref{fig:boxc}. For $\Gamma = 0.01\delta$ and $N_d=201 \gg \delta/\Gamma$, we are visibly approximating the boxcar function, up to small edge errors.

\section{Boxcar transmission function: current, noise and precision}
\label{app:E}
We consider a boxcar transmission function with width $2\delta$ shifted in the positive direction by an amount $\theta$. The transmission function can be expressed as 
\begin{align}
    \mathcal T^{\text{box}}\left( \epsilon\right) = \Theta\left(\epsilon+\delta - \theta \right) - \Theta\left(\epsilon-\delta-\theta \right),
\end{align}where $\Theta(\epsilon)$ is the Heaviside function. Using \eqref{eq:J} we find that 
\begin{equation}
\begin{aligned}
    \expval{I} = \frac{2e}{h}k_BT \left(-\log \left(e^{\frac{e V+\epsilon_F}{k_BT}}+e^{\frac{\delta +\theta }{k_BT}}\right)+\log \left(e^{-\frac{\delta }{k_BT}} \left(e^{\frac{\delta +e V+\epsilon_F}{k_BT}}+e^{\frac{\theta }{k_BT}}\right)\right)-\log \left(e^{\frac{\theta -\delta }{k_BT}}+e^{\frac{\epsilon_F}{k_BT}}\right)+\log \left(e^{\frac{\delta +\theta }{k_BT}}+e^{\frac{\epsilon_F}{k_BT}}\right)\right) 
\end{aligned}
\end{equation}

The fluctuations can similarly be calculated,
\begin{equation}
    \begin{aligned}
        \left\langle\left\langle I^2 \right\rangle\right\rangle &= 
        \int d\epsilon \,\left[\Pi_\theta\left(\epsilon\right) \left(1-\Pi_\theta\left(\epsilon\right) \right) \mathcal F_1\left(\epsilon\right) + \Pi_\theta\left(\epsilon\right) \mathcal F_2\left(\epsilon\right)\right]=\int d\epsilon\, \Pi_\theta\left(\epsilon\right) \mathcal F_2\left(\epsilon\right) \\
        &= 4 \frac{e^2}{h} k_BT \sinh \left(\frac{\delta }{k_BT}\right) \left(\frac{1}{\cosh \left(\frac{e V-\theta +\epsilon_F}{k_BT}\right)+\cosh \left(\frac{\delta }{k_BT}\right)}+\frac{1}{\cosh \left(\frac{\delta }{k_BT}\right)+\cosh \left(\frac{\theta -\epsilon_F}{k_BT}\right)}\right)
    \end{aligned}
\end{equation}The precision rate can be evaluated using Eq. (1) in the main text,
\begin{align}
    \gamma^\text{box} = \frac{ \frac{4}{hk_BT}\sinh \left(\frac{\delta }{k_BT}\right) \sinh ^2\left(\frac{e V}{2 k_BT}\right) \sinh ^2\left(\frac{e V-2 \theta +2 \epsilon_F}{2 k_BT}\right)}{\left(\cosh \left(\frac{\delta }{k_BT}\right)+\cosh \left(\frac{\theta -\epsilon_F}{k_BT}\right)\right) \left(\cosh \left(\frac{e V-\theta +\epsilon_F}{k_BT}\right)+\cosh \left(\frac{\delta }{k_BT}\right)\right) \left(\cosh \left(\frac{e V-\theta +\epsilon_F}{k_BT}\right)+2 \cosh \left(\frac{\delta }{k_BT}\right)+\cosh \left(\frac{\theta -\epsilon_F}{k_BT}\right)\right)}
\end{align}

\subsection{Linear response}

\subsubsection{$k_BT\gg eV$}
For $k_BT\gg eV$, the linear-response expressions are given below, 
\begin{align}
\expval{I}_{\text{box}}^{\text{LR}} = 2\frac{e^2V}{h}\frac{\sinh \left(\frac{\delta }{k_BT}\right)}{\cosh \left(\frac{\delta }{k_BT}\right)+\cosh \left(\frac{\theta -\epsilon_F }{k_BT}\right)}
\end{align}and 
\begin{align}
    \left\langle\left\langle I^2\right\rangle\right\rangle^{\text{box}}_{\text{LR}} = 4k_BT \frac{e^2}{h}\frac{ \sinh \left(\frac{\delta }{k_BT}\right)}{\cosh \left(\frac{\delta }{k_BT}\right)+\cosh \left(\frac{\theta -\epsilon_F }{k_BT}\right)}.
\end{align}Therefore,
\begin{align}
    \gamma^{\text{box}}_{\text{LR}} = \frac{e^2V^2 \sinh \left(\frac{\delta }{k_BT}\right) \sinh ^2\left(\frac{\theta -\epsilon_F }{k_BT}\right)}{2 h(k_BT)^3 \left(\cosh \left(\frac{\delta }{k_BT}\right)+\cosh \left(\frac{\theta -\epsilon_F }{k_BT}\right)\right)^3}
\end{align}

\subsubsection{$T\to0$}
Here, we consider the zero-temperature limit for the boxcar transmission function. Depending on where the boxcar function lies with respect to the bias window, different limits can be obtained. In particular, we would like $(\partial_\theta \expval{J})^2$ to be as large  as possible and $\left\langle\left\langle I^2\right\rangle\right\rangle$ to be as small as possible. However, if the bias window lies within the boxcar shape, then the current takes the form
\begin{align}
\lim_{T\to0}\expval{I}^{\text{box}} = 2\frac{e}{h}\int_{\epsilon_F}^{\epsilon_F+eV} d\epsilon = 2\frac{e^2V}{h}= G_0V,
\end{align}which is independent of $\theta$, and so the derivative $\partial_\theta \expval{J}=0$. In addition, the boxcar cannot lie outside the bias window because in such a case, the zero-temperature limit of the current will be zero. Consider the case when the boxcar lies within the bias window. In this case, 
\begin{align}
\lim_{T\to0}\expval{I}^{\text{box}} = 2\frac{e}{h}\int_{-\delta+\theta}^{\delta+\theta} d\epsilon = 4\frac{e}{h}\delta,
\end{align}which is also independent of $\theta$, and the derivative is again zero. This means that this case is not useful. \par
Consider now the transmission function to lie partially within the bias window. Here, the current takes the form
\begin{equation}
\begin{aligned}
\lim_{T\to0}\expval{I}^{\text{box}} & = 2\frac{e}{h}\int_{-\delta+\theta}^{\epsilon_F+eV}d\epsilon,\,\,\,\,\text{or}\,\,\,\,2\frac{e}{h}\int^{\delta+\theta}_{\epsilon_F}d\epsilon
\\&= 2\frac{e}{h}\left[\left(\epsilon_F +eV\right) - \left(-\delta+\theta \right)\right],\,\,\,\,\text{or}\,\,\,\,2\frac{e}{h}\left[\left(\delta+\theta\right) - \epsilon_F\right],
\end{aligned}
\end{equation}depending on where the transmission function lies with respect to the bias window. Therefore the derivative is $\partial_\theta \expval{I} =\pm 2e/h$. The fluctuations in this case take the form
\begin{equation}
\begin{aligned}
\lim_{T\to0}\left\langle\left\langle I^2\right\rangle\right\rangle^{\text{box}} &= \lim_{T\to0} 4\frac{e^2}{h}\int_{-\delta+\theta}^{\epsilon_F+eV} d\epsilon\, \mathcal F_2\left( \epsilon\right),\,\,\,\,\text{or}\,\,\,\, \lim_{T\to0} 4\frac{e^2}{h}\int_{\epsilon_F}^{\delta+\theta} d\epsilon \,\mathcal F_2\left( \epsilon\right)
\\& = 0,
\end{aligned}
\end{equation}because $\lim_{T\to0}\mathcal F_2=0$. We therefore obtain a diverging precision rate at zero temperature.

\end{document}